V. F. Kovalenko[1], Dr.Sci., Prof.,
M. V. Petrychuk[1], Ph.D.,
B. M. Tanygin[1**], Ph.D.,
S. I. Shulyma[1,*], post grad. student.

# Ferrofluid aggregates phase transitions in the planar magnetic field

The influence of the cyclic heating and cooling on properties of the aggregates (aka "ferrofluid clusters") in a ferrofluid, which made on the basis of magnetite nanoparticles, are investigated. The heating of the ferrofluid layer with such aggregates leads to equalization of the concentration between high- and low-concentrated phases. The temperature of the equalization of the phase concentrations was determined at different values of an external constant magnetic field, which was applied parallel to the layer of the ferrofluid. The temperature of the destruction of a periodic structure of the magnetic aggregates, which were formed during cooling of a homogeneous phase of the ferrofluid, was obtained at the different values of the applied external magnetic field.

[1] Taras Shevchenko National University of Kyiv, 03022, Kyiv, Glushkova st., 4g

[*]E-mail: kiw_88@mail.ru, [**]E-mail: b.m.tanygin@gmail.com

## Introduction

Ferrofluids (FF) are colloidal suspensions of the magnetic nanoparticles in the carrier liquid. Latest year's increasing interest to the FF is related to the ability of their usage for different applied purposes: *in vivo* drug deliveries [2], hyperthermia [3], as an agent of the magneto-resonance tomography [4], others [5,6].

Based on the internal structure, FF structure can be systemized as: (i) liquid droplet aggregates (LDAs) [7], and (ii) solid aggregates [8]. The LDA FF research interest relates to wide application of aerosols in household and technics. Another reason is a fundamental interest to obtain more complete knowledge about the ferrofluid. The LDA FF behavior is well researched in the Ref. [1]. However, research work of cyclic heating and cooling of such FF in the presence of external magnetic field was not done. According to Ref. [9] process of the cooling of previously heated FF layer leads to emergence of the self-organized structure of the LDA. Here, normal (to FF layer) magnetic field was applied. Investigation of large LDA (LLDA) (volume $V_c = 0.06\text{-}0.13 \text{ mm}^3$) was not performed as well.

Thus, aim of the present research is an experimental investigation of the cyclic heating/cooling FF containing LLDA in a planar magnetic field.

## Experiment

In scope of our research, we used FF "magnetite in kerosene". An oleic acid was used as an anticoagulant. Initial volume concentration of the FF dispersed phase was equal to 12%. The FF was carried by $100 \mu$ container bounded between substrate and electro-conductive cover slip. Heating of the FF layer had been performing by electrical current. Temperature was measured by thermal sensor DS1820.

The LLDA formation have been forced by magnetic field ~1 kOe, created by permanent magnet contacted the FF layer boundary surface (Fig.1a).

External planar magnetic field was applied by Helmholtz coil connected to the DC power supply. Changes of the current magnitude correspond to changes of this magnetic field in a range: 0 .. 1.5 kOe. This value was measured by the magnetic field sensor SS490. Inhomogeneity of the magnetic field in the place of sensor is estimated as < 2%.

By means of usage of the CMOS camera (connected to personal computer) and further digital analysis, the visual observation of sizes and structure of the (L)LDA was performed.

## Results & Discussion

### *Ferrofluid heating in the zero external field*

It was shown, that first LLDA FF heating leads to the destruction of the LLDA structure, meaning by setting equal concentrations between high and low concentrated FF phases observed simultaneously. (Fig. 1b). At the temperature T=52-56°C, the LLDA fully transforms to the homogeneous FF phase, which is related to domination of the Brownian motion upon the LLDA surface tension energy.

Further (first) FF cooling leads to emergence of the quasi-periodical structure of small sphere-like shape LDAs (Fig.2 a,b). The second heating of the FF layer leads to return to the equal concentrations phases (same as first heating) (Fig.3). The second cooling of this FF layer leads to emergence of the same sphere-like shape LDAs structure (Fig.2).

### *Applying of the non-zero external field to large liquid droplet aggregate*

Applying of the external magnetic field transforms the LLDA into a quasi-periodical structure of the rod-shaped aggregates (RSAs) aligned to the field direction. They can be either connected or not (Fig.4). The RSA length increases at the field magnitude (Fig.4). At the magnetic field magnitude <100 Oe, the RSAs were not split. They were connected by cross connection perpendicular to the magnetic field direction (Fig.4a). At the magnetic field magnitude >100 Oe, the LLDA was transformed into a quasi-periodic structure of the separate RSAs (Fig.4c). We relate this phenomenon with assumption, that H<100 Oe corresponds to higher energies of the deformed LLDA compare to magnetostatic energy of the separated RSAs system.

### *Ferrofluid heating in the non-zero external field*

In scope of the present research work, we made 5 cycles of measurements at the fixed external magnetic field. Each cycle is a two consecutive heating/cooling pairs. Increase of temperature in the FF layer leads to increase of RSAs length (both connected and split ones). Connected RSAs were fully separated at the H=50 Oe and T=30-36 °C (Fig.5). Thus, increase of the temperature leads to increase of the heating oscillation intensity increase and T=30-36 °C corresponds to respective destruction of the cross connections.

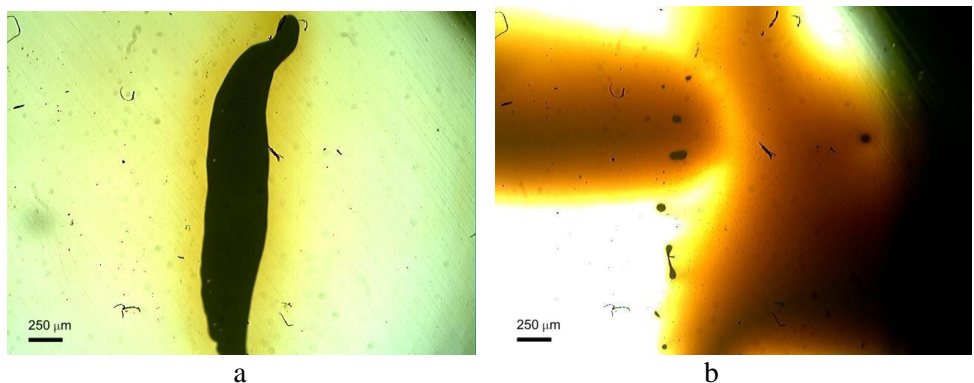

Fig1. Transformation of the large liquid-droplet aggregate at the ferrofluid heating to the following temperatures: T=26 °C (a); T=55 °C (b)

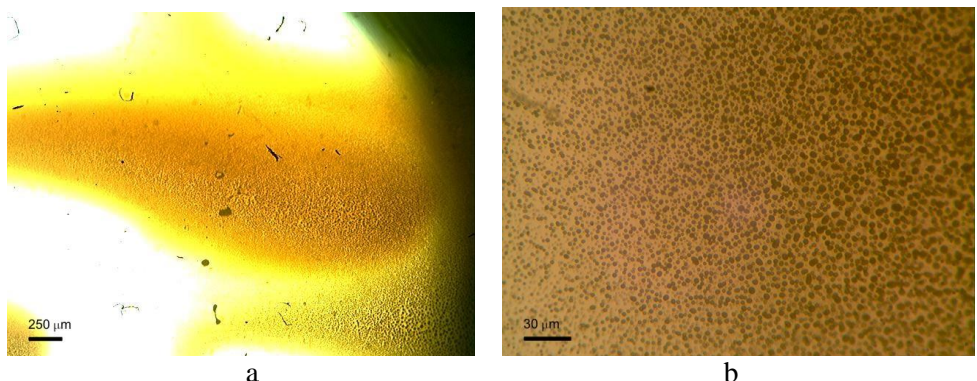

Fig.2. Quasi-periodical structure of the liquid-droplet aggregates at the fixed temperatures температурі T=45 °C (different scales: a and b)

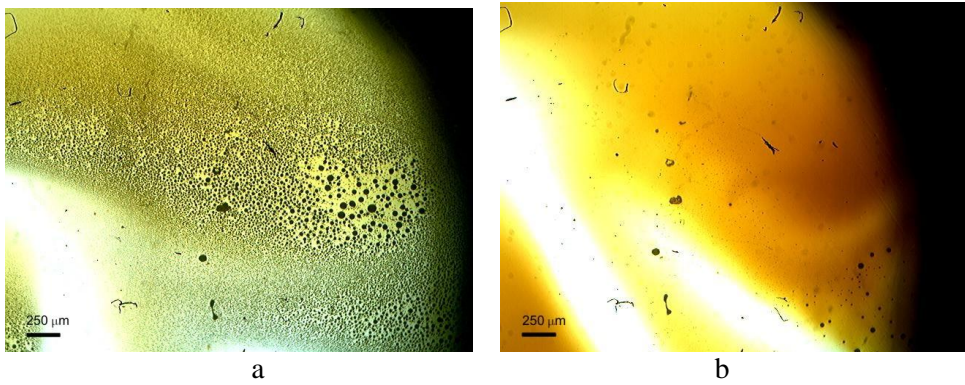

Fig.3. Transformation of the quasi-periodical structure of the liquid-droplet aggregates at the second heating: T=26 °C (a); T=56 °C (b)

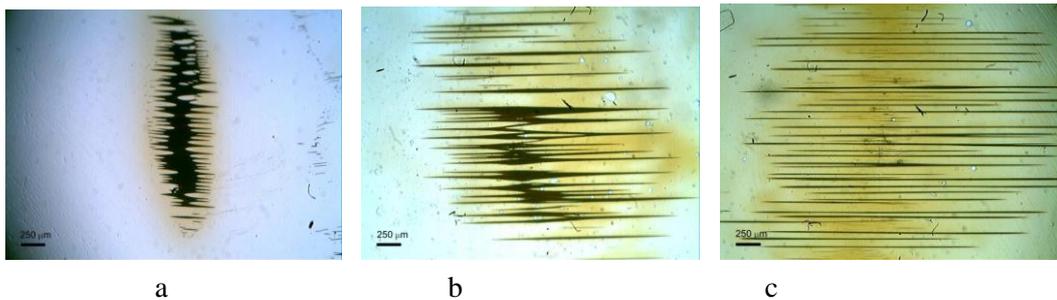

Fig.4. Transformation of the quasi-periodical structure of the rod-shaped aggregates in external magnetic field (T=26°C): H=30 Oe (a); H=100 Oe (b); H=1300 Oe (c).

In case of the ferrofluid with different initial concentration of the dispersed phase, this temperature can be different. Here, the RSA structure remains unchanged. The first FF heating leads to destruction of the RSA structure and stabilization of homogeneous phase, similar to one presented on Fig.1b. The temperature of the RSA destruction depends on the external magnetic field magnitude (Fig.6) in a range T=52..54°C. The tangent of the linear approximation (Fig.6) is equal to ~1.1·10⁻³ °C/Oe.

First cooling of this homogenous phase in an external magnetic field leads to creation of the similar quasi-periodical structure.

Comparing to zero-field case (Fig.2) this is a structure of the RSAs (Fig.7).

Second FF heating in a non-zero field leads to destruction of the RSAs structure, though respective temperature is different compare to the first heating (Fig.8). The tangent of the linear approximation (Fig.8) is equal to ~ −3.4·10⁻³ °C/Oe. Thus, effect of the external field is opposite compare to the first heating.

### Discussion

In order to explain former phenomenon (opposite impact of field on the destruction temperatures) we are suggesting the following model, related to changes of the internal arrangement of the nanoparticles inside aggregates.

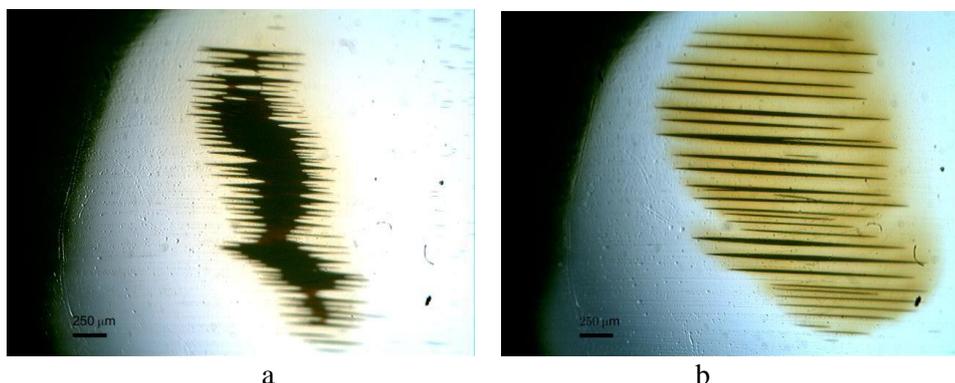

Fig.5. Changes of the periodic structure of the rod-shaped aggregates at the H=50 Oe: heating of the ferrofluid layer from room temperature T=26 °C (a) to T=35°C (b)

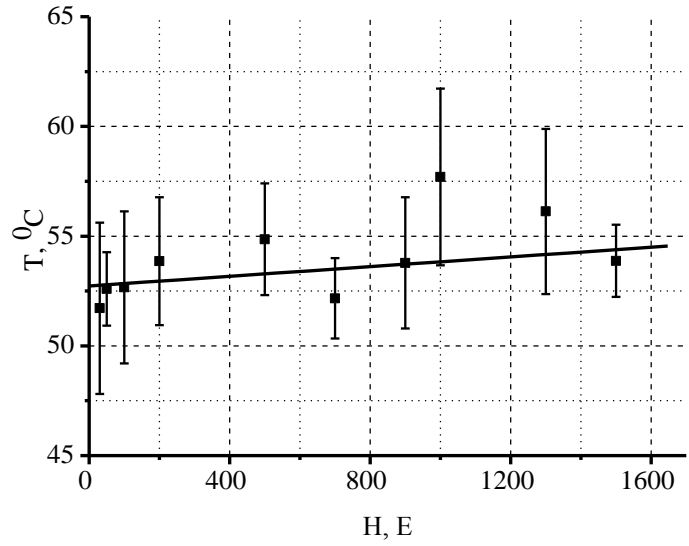

Fig.6. Dependence of the rod-shaped aggregates destruction temperature on the field magnitude (first heating).

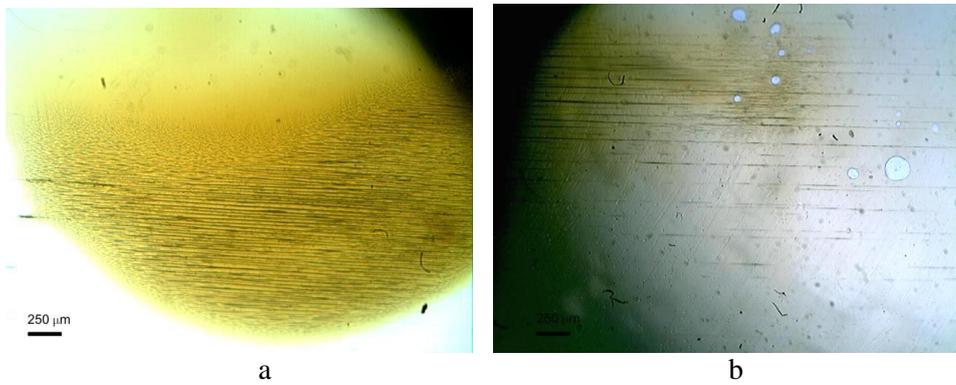

<div align="center">a         b</div>

Fig.7. Rod-shaped aggregates quasi-periodic structure (T=40 °C) emerged after homogeneous phase cooling in the external magnetic field: H=50 Oe (a); H=1300 Oe (b)

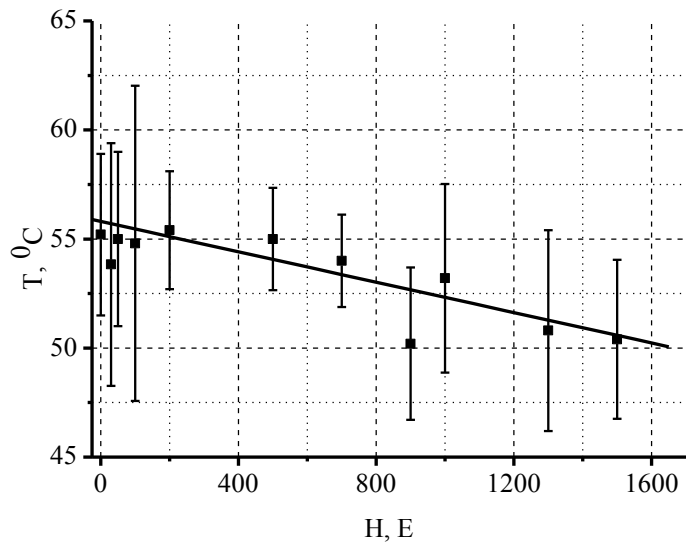

Fig.8. Dependence of the rod-shaped aggregates destruction temperature on the field magnitude (second heating).

Initially, LLDA consists of the primary [10] aggregates one-dimensional chains arranged into the vortex-like structure (confined magnetic flux). This case is a similar one to the primary aggregate internal structure [10]. After first FF cooling in the non-zero field, this vortex-like structure transforms into structure of weakly interacted primary aggregates one-dimensional chains, which start to be sensitive to impact of the applied field and temperature.

Additionally, the increase of the magnetic field magnitude leads to the RSAs split, which also reduce temperature of the emergence of the homogeneous phase.

## Conclusions

Thus, in scope of the present work, the behavior of the large liquid droplet aggregates in the applied magnetic field and under the cyclic heating and cooling was investigated. The following conclusions were made:

1. Heating of the ferrofluid with large liquid-droplet aggregates or rod-shaped aggregates leads to destruction of their structure, which was related to domination of the thermal energy upon the surface tension of an aggregate.

2. Initial (before heating) structure of the liquid-droplet aggregate is fully destroyed in a range T=52..56°C regardless of the magnitude of the applied magnetic field. The consecutive phase is a homogeneous ferrofluid phase without visible liquid-droplet aggregates.

3. Cooling of the homogeneous ferrofluid in non-zero or zero magnetic field leads to creations of the structure of rod-shaped or spherical liquid-droplet aggregates respectively.

4. Second heating of the ferrofluid in the non-zero magnetic field leads to destruction of the rod-shaped aggregates. Respective temperature reduces at the increase of the magnetic field magnitude. In our opinion, such behavior relates to the changes of the internal nanoparticles arrangement inside the aggregate.